%% file: QM18Proceedings_HarryAndrews_v4.tex
\documentclass[3p,times,procedia]{elsarticle}
\usepackage{nupha_ecrc}
\usepackage{caption}
\usepackage{subcaption}
\input{Include/Include}

\volume{00}

\firstpage{1}

\journalname{Nuclear Physics A}

\runauth{}

\jid{nupha}

\jnltitlelogo{Nuclear Physics A}

\usepackage{amssymb}
\usepackage[figuresright]{rotating}

\begin{document}

\begin{frontmatter}

%% Title, authors and addresses

%% use the tnoteref command within \title for footnotes;
%% use the tnotetext command for the associated footnote;
%% use the fnref command within \author or \address for footnotes;
%% use the fntext command for the associated footnote;
%% use the corref command within \author for corresponding author footnotes;
%% use the cortext command for the associated footnote;
%% use the ead command for the email address,
%% and the form \ead[url] for the home page:
%%\title{Title\tnoteref{label1}}
%% \tnotetext[label1]{}
%% \author{Name\corref{cor1}\fnref{label2}}
%% \ead{email address}
%% \ead[url]{home page}
%% \fntext[label2]{}
%% \cortext[cor1]{}
%% \address{Address\fnref{label3}}
%% \fntext[label3]{}

%% Instructions from Editor: Please use the following \dochead only in the preprint version (e-print arXiv etc.); 
%% use empty \dochead{} when submitting to Nuclear Physics A!
\dochead{XXVIIth International Conference on Ultrarelativistic Nucleus-Nucleus Collisions\\ (Quark Matter 2018)}
%\dochead{}
%% Use \dochead if there is an article header, e.g. \dochead{Short communication}
%% \dochead can also be used to include a conference title, if directed by the editors
%% e.g. \dochead{17th International Conference on Dynamical Processes in Excited States of Solids}

\title{Exploring the Phase Space of Jet Splittings at ALICE using Grooming and Recursive Techniques}

%% use optional labels to link authors explicitly to addresses:
%% \author[label1,label2]{<author name>}
%% \address[label1]{<address>}
%% \address[label2]{<address>}

\author{Harry Arthur Andrews \\ on behalf of the ALICE collaboration}

\address{University of Birmingham}

\begin{abstract}
Hard splittings in the evolution of a jet may be modified by the presence of a dense strongly interacting medium. Grooming procedures can be used to isolate such hard components of a jet and allows one to focus on the two subjets resulting from a sufficiently hard partonic splitting. Measurements of the symmetry parameter (\zg), angular separation (\Rg) and number of splittings (\nsd) of Soft Drop groomed jets are reported as measured with the ALICE Detector in pp and Pb--Pb collisions at \sqrts = 7 TeV and \sqrtsNN = 2.76 TeV respectively. The use of recursive splittings and their mappings to identify interesting regions of phase space are also discussed.
\end{abstract}

\begin{keyword}
Jets, Jet Substructure, Heavy-Ion Collisions, QCD, Quark-Gluon Plasma
%% keywords here, in the form: keyword \sep keyword

%% MSC codes here, in the form: \MSC code \sep code
%% or \MSC[2008] code \sep code (2000 is the default)

\end{keyword}

\end{frontmatter}

%%
%% Start line numbering here if you want
%%
% \linenumbers

%% main text
\section{Introduction}
\subsection{Jet Grooming}
\label{sect:Intro}
In relativistic heavy ion collisions, exchanges of large momentum between partonic constituents occur in the early stages and subsequently the scattered partons traverse the hot and dense medium of deconfined colour charges produced. Jets, that evolve from these scattered partons, in the presence of this medium are known to experience energy loss, commonly referred to as jet quenching~\cite{Armesto:2011ht}. %Jet quenching has been observed and studied extensively in heavy ion collisions at both RHIC (BNL)~\cite{Betz:2012hv} %~\cite{Adcox:2001jp}~\cite{Adams:2003kv}~\cite{Adams:2003im}~\cite{Adler:2002xw}~\cite{Adler:2002tq}%
% and the LHC (CERN)~\cite{Spousta:2013aaa}%~\cite{Aad:2012vca}~\cite{Aad:2014bxa}~\cite{CMS:2012aa}~\cite{Chatrchyan:2011sx}~\cite{Chatrchyan:2012nia}~\cite{Chatrchyan:2012gt}~\cite{Chatrchyan:2014gia}~\cite{CMS:2012rba}~\cite{Aamodt:2010jd}~\cite{Abelev:2012hxa}~\cite{Abelev:2013kqa}%.
 This energy loss is thought to occur dominantly due to gluon radiation induced by the medium relative to vacuum-like evolution~\cite{Baier:1994tc}\cite{Gyulassy:1990}. Understanding how the fragmentation pattern of jets is modified by the medium is the main scope of this analysis. The soft components of these modifications are very difficult to measure and model, however, studying the limit of the hardest components of the parton shower, that can occur in cone, can provide information on how these medium-induced processes take place. In addition to studying medium-induced fragmentation, investigating the properties of the the hard components of the parton shower can help to elucidate the role of coherent and decoherent emission in the presence of the coloured medium~\cite{Mehtar-Tani:2016aco}. 
\newline \indent
 In order to study these leading components of the parton shower, jets are groomed to identify hard splittings using the Soft Drop jet grooming algorithm~\cite{Dasgupta:2013ihk}\cite{Larkoski:2014wba}. Soft Drop begins the grooming by first reclustering the jets with a given algorithm, most commonly Cambridge-Aachen (CA)~\cite{FastJetCA}, and unwinding the clustering one step. The resulting pair of subjets are then considered and their momentum fraction, $z$, is calculated as $z = \frac{min(p_{\rm{T,1}},p_{\rm{T,2}})}{p_{\rm{T,1}}+p_{\rm{T,2}}}$, where $\it{p}_{\rm{T,1}}$ and $\it{p}_{\rm{T,2}}$ are the transverse momenta of the two subjets. If this momentum fraction satisfies the grooming condition: 
 
 %\begin{equation}
%z = \frac{min(p_{\rm{T,1}},p_{\rm{T,2}})}{p_{\rm{T,1}}+p_{\rm{T,2}}}, \\
%\label{eq:momfraction}
%\end{equation}
%\noindent where $\it{p}_{\rm{T,1}}$ and $\it{p}_{\rm{T,2}}$ are the transverse momenta of the two subjets. If this momentum fraction satisfies the grooming condition:

\begin{equation}
z > z_{\rm{cut}} ~ \Big( \frac{\DeltaR}{R_{0}} \Big)^{\beta} \\
\label{eq:softdrop}
\end{equation}
\noindent
then the splitting identified is considered sufficiently hard and the grooming procedure is stopped. If the condition is not satisfied then the softer subjet is discarded and the clustering of the harder branch is unwound an additional step; this process is repeated until a splitting satisfying (\ref{eq:softdrop}) is found. The momentum fraction at this stage is identified as the groomed momentum fraction $\it{z}_{\rm{g}}$. The angular separation of the two subjets, as defined in $\eta-\varphi$ space, is another important parameter of the splitting and is assigned as the groomed radius $R_{\rm{g}}=\sqrt{(\eta_{\rm{subjet,1}}-\eta_{\rm{subjet,2}})^{2}+(\varphi_{\rm{subjet,1}}-\varphi_{\rm{subjet,2}})^{2}}.$ Reclustering with the CA algorithm is designed to replicate the angular ordering of QCD vacuum splittings and, to leading order, the measurement of $\it{z}_{\rm{g}}$ in vacuum reproduces the Altarelli-Parisi splitting functions. 

%\noindent then the splitting identified is considered sufficiently hard and the grooming procedure is stopped. If the condition is not satisfied then the softer subjet is discarded and the clustering of the harder branch is unwound an additional step; this process is repeated until a splitting satisfying (\ref{eq:softdrop}) is found. The momentum fraction at this stage is identified as the groomed momentum fraction $\it{z}_{\rm{g}}$. The angular separation of the two subjets, as defined in $\eta-\varphi$ space, is another important parameter of the splitting and is assigned as the groomed radius $\it{R}_{\rm{g}}$:
%\begin{equation}
%R_{\rm{g}}=\sqrt{(\eta_{\rm{subjet,1}}-\eta_{\rm{subjet,2}})^{2}+(\varphi_{\rm{subjet,1}}-\varphi_{\rm{subjet,2}})^{2}}. \\
%\label{eq:DeltaRSubJet}
%\end{equation}
 %Reclustering with the CA algorithm is designed to replicate the angular ordering of QCD vacuum splittings and, to leading order, the measurement of $\it{z}_{\rm{g}}$ in vacuum reproduces the Altarelli-Parisi splitting functions. 
% \newline \indent
In addition to studying the parameters of the leading hard splitting of jets the number of them that arise in the evolution of the jet can help identify any additional splittings occurring due to the presence of the medium. To count the number of splittings, the grooming is continued past the first splitting to satisfy (\ref{eq:softdrop}) and, following the hardest branch at each stage, the total number that pass the condition are counted and assigned as $\it{n}_{\rm{SD}}$.
%%%~~~~~~~~~~~~~~~~~~~~~~~~~~~~~~~~~~~~~~~~%%%
\subsection{The Lund Plane}
	
\begin{figure}[!h]
\centering
\includegraphics[width=0.4\textwidth]{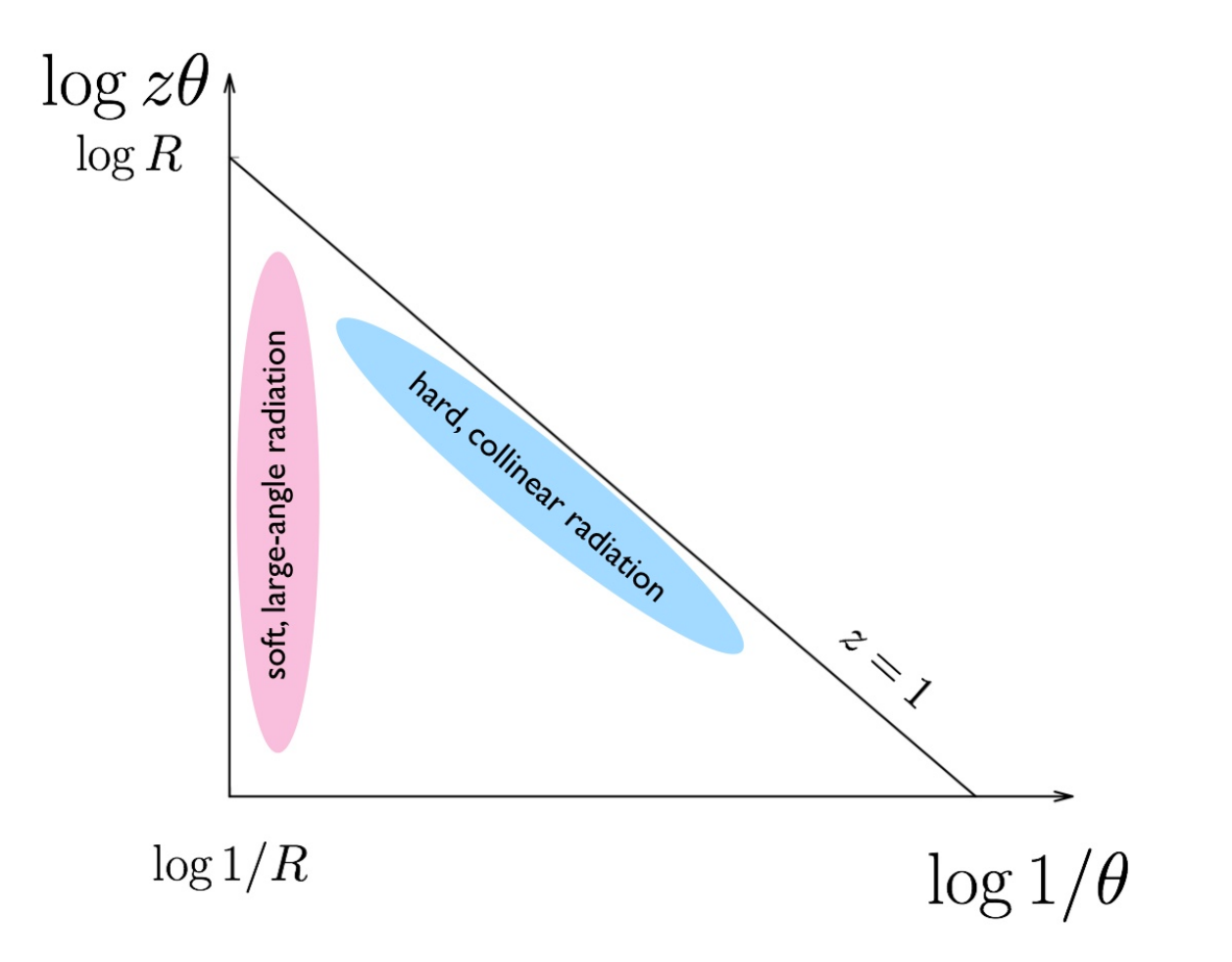}
\includegraphics[width=0.35\textwidth]{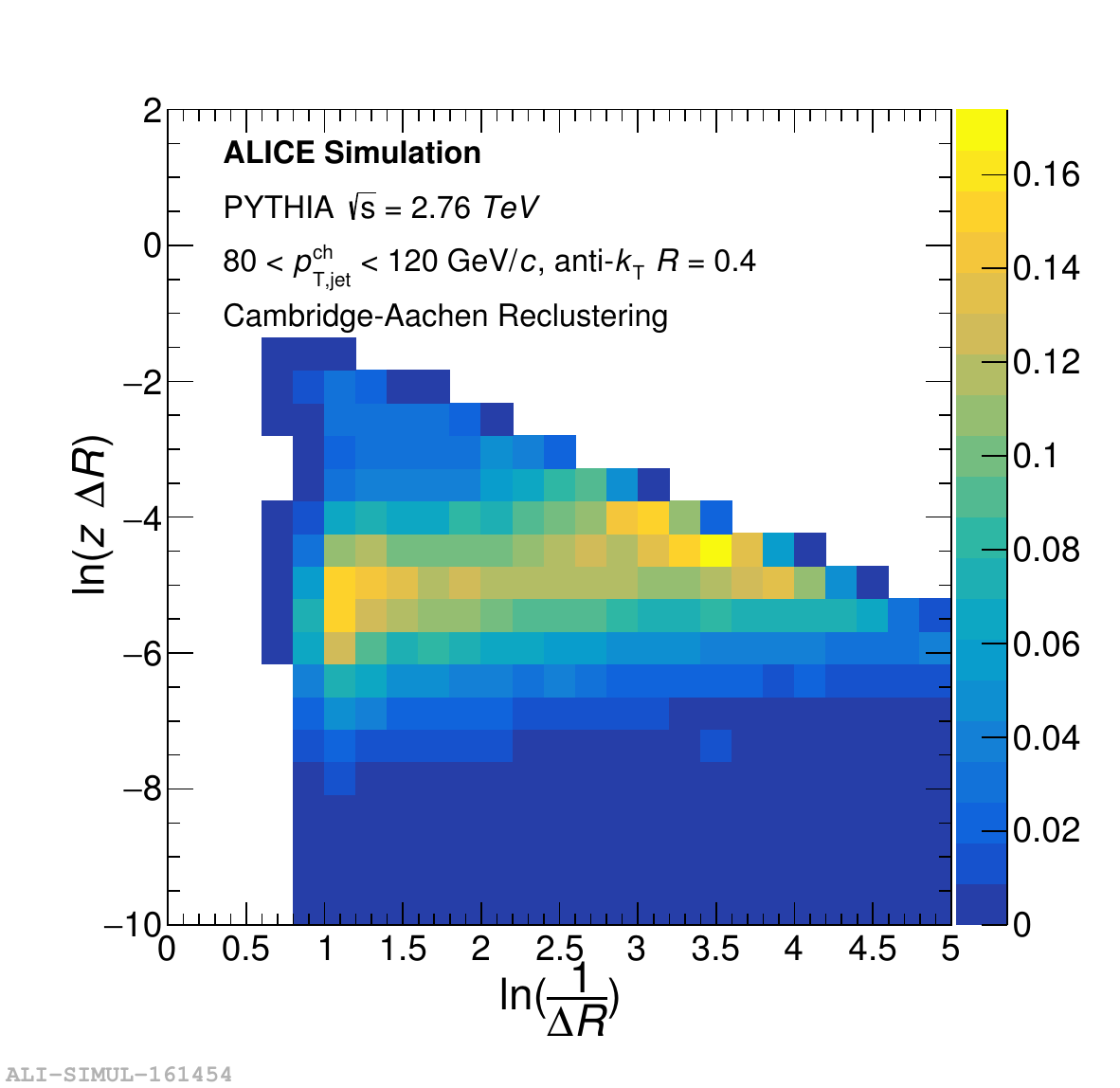}

\caption{Lund kinematical diagram representation of splittings with limits imposed by a jet resolution $R$ (left)~\cite{THInst}, and populated for splittings in vacuum PYTHIA 6 Perugia 11 (right).}
\label{fig:LundKonrad}
\end{figure}

A very useful representation of the splittings is the Lund kinematical diagram. The Lund diagram represents the 1$\rightarrow$ 2 splitting process of a parton on the two axes as shown in Fig.\ref{fig:LundKonrad}. The axes reflect the gluon emission probability given by:
\begin{equation}
dP = 2\frac{\alpha_{s}C_{i}}{\pi}\rm{dlog}(z\theta)\rm{dlog}\frac{1}{\theta},
\label{eq:gluonprob}
\end{equation}

\noindent where $\theta$ is the aperture angle of the splitting and $C_{i}$ is the colour factor for a gluon radiated off an initial quark ($C_{i} = C_{F}$) or gluon ($C_{i} = N_{c}$).
\newline \indent
Representing the phase space of splittings in this way allows one to isolate regions where different medium-induced mechanisms are expected to contribute to the modification of the parton shower splitting function. As it can be seen in Fig.~\ref{fig:LundKonrad} (left) the region where soft wide angle splittings dominate can be clearly separated from the region of hard collinear splittings. Fig.~\ref{fig:LundKonrad} (right) shows how this diagram is populated with recursive splittings generated using PYTHIA and identified using CA reclustering. 

\section{Analysis Procedure}

In this analysis two collisions systems were considered, pp and Pb--Pb. The pp collisions were measured by the ALICE detector at \sqrts = 7 TeV and the Pb--Pb collisions at \sqrtsNN = 2.76 TeV. Charged jets were reconstructed with a minimum track transverse momentum of $p_{\rm{T,track}}^{\rm{ch}} > $150 MeV and clustered using the FastJet anti-$k_{\rm{T}}$ algorithm~\cite{FastJetAntikt} with a resolution $R =$ 0.4~\cite{Cacciari:2011ma}. The jet track candidates were combined using the E-scheme prescription. In the pp measurements there was no background subtraction applied to the jet and in the Pb--Pb measurements constituent subtraction was used~\cite{Berta:2014eza}. 

\section{Results}
\begin{figure}[!h]
\centering
\includegraphics[width=0.45\textwidth]{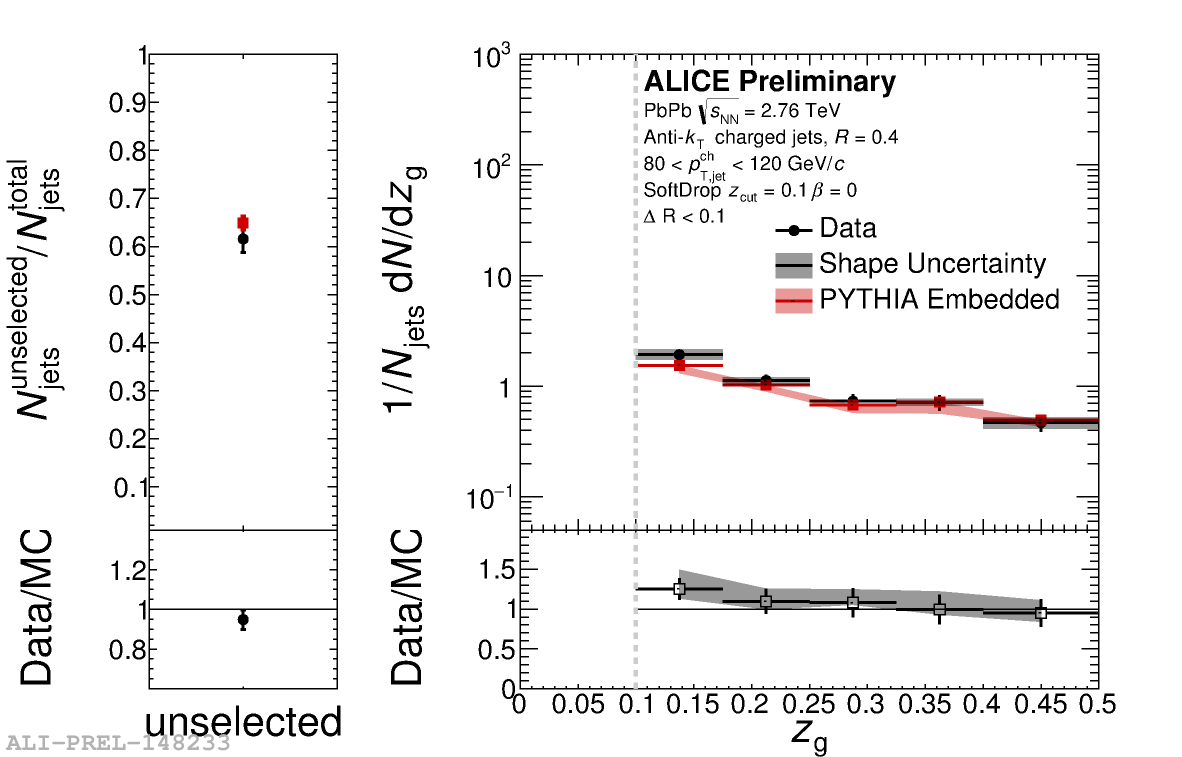}
\includegraphics[width=0.45\textwidth]{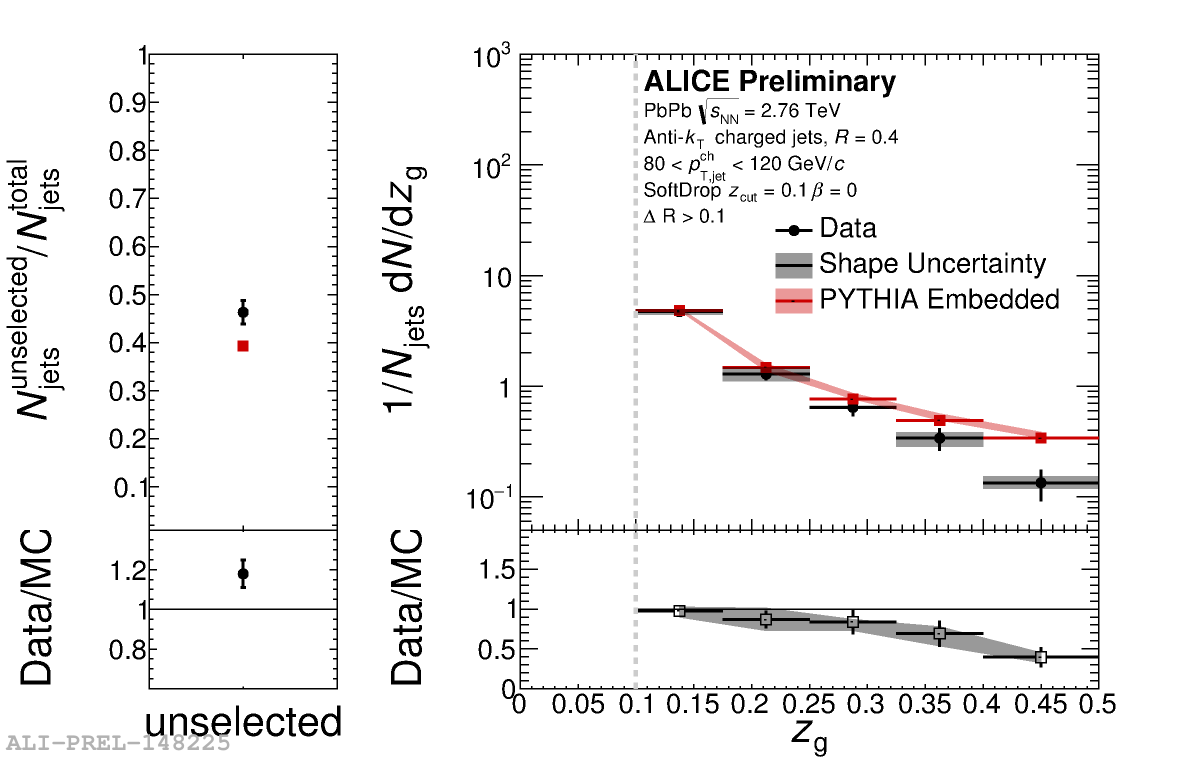}

\caption{Raw Pb--Pb distributions of $z_{\rm{g}}$ for $R=0.4$ jets with varying minimum angular separation of subjets ($R_{\rm{g}}$) for jets with $ 80 \leq \pTjetchrec < 120$ GeV/$c$. The distributions are normalised to the total number of jets in the sample in this \pTjetchrec ~bin without cuts on $R_{\rm{g}}$.}
\label{fig:PbPbrawzg}
\end{figure}

The measurements of \zg ,~\Rg ~and \nsd ~in minimum bias pp collisions were made and fully corrected using a 2D-Bayesian unfolding approach to correct for detector effects. The detector effects were characterised using GEANT3 on PYTHIA generated jets. The results for the pp collisions are presented in the range $40 \leq \pTjetch <$ 60 GeV/$c$~\cite{QM18Andrews}. Due to a lack of pp statistics at the Pb--Pb collision energy of $\sqrtsNN ~= 2.76~ \rm{TeV}$ and the good agreement observed between pp data and PYTHIA, PYTHIA is used as the vacuum reference in the Pb--Pb analysis.
\newline \indent
For Pb--Pb collisions the measurements of \zg ~and \nsd ~were made in the range 80 $\leq \pTjetchrec <$ 120 GeV/$c$ to avoid contamination from combinatorial background and compared to a smeared reference generated by embedding PYTHIA jets into 0-10\% central Pb--Pb events. The measurement of \zg ~is studied differentially in angular separation (\Rg) and as shown in Fig.~\ref{fig:PbPbrawzg}, this has a significant effect on the distribution. For splittings at large angles (right) a steepening of the distribution is observed as well as an overall suppression in the number of jets selected by the grooming procedure above this threshold. Conversely in the collinear limit (left) no such modification is observed and the ratio of jets selected in this region is consistent between the Pb--Pb data and the vacuum reference. It is important to note that the distributions are normalised by the total number of jets in the given transverse momentum bin. This means that integral of the distributions in the range 0.1$ \leq z_{\rm{g}} < $0.5 is not constant between the distributions and the bin labelled ''unselected'' shows the fraction of jets that do not enter this part of the distribution. Doing the normalisation this way allows one to correctly attribute any modification of the distribution to a suppression or enhancement.

\begin{figure}[!h]
\centering
\begin{minipage}{.45\textwidth}
\centering
\includegraphics[width=0.9\textwidth]{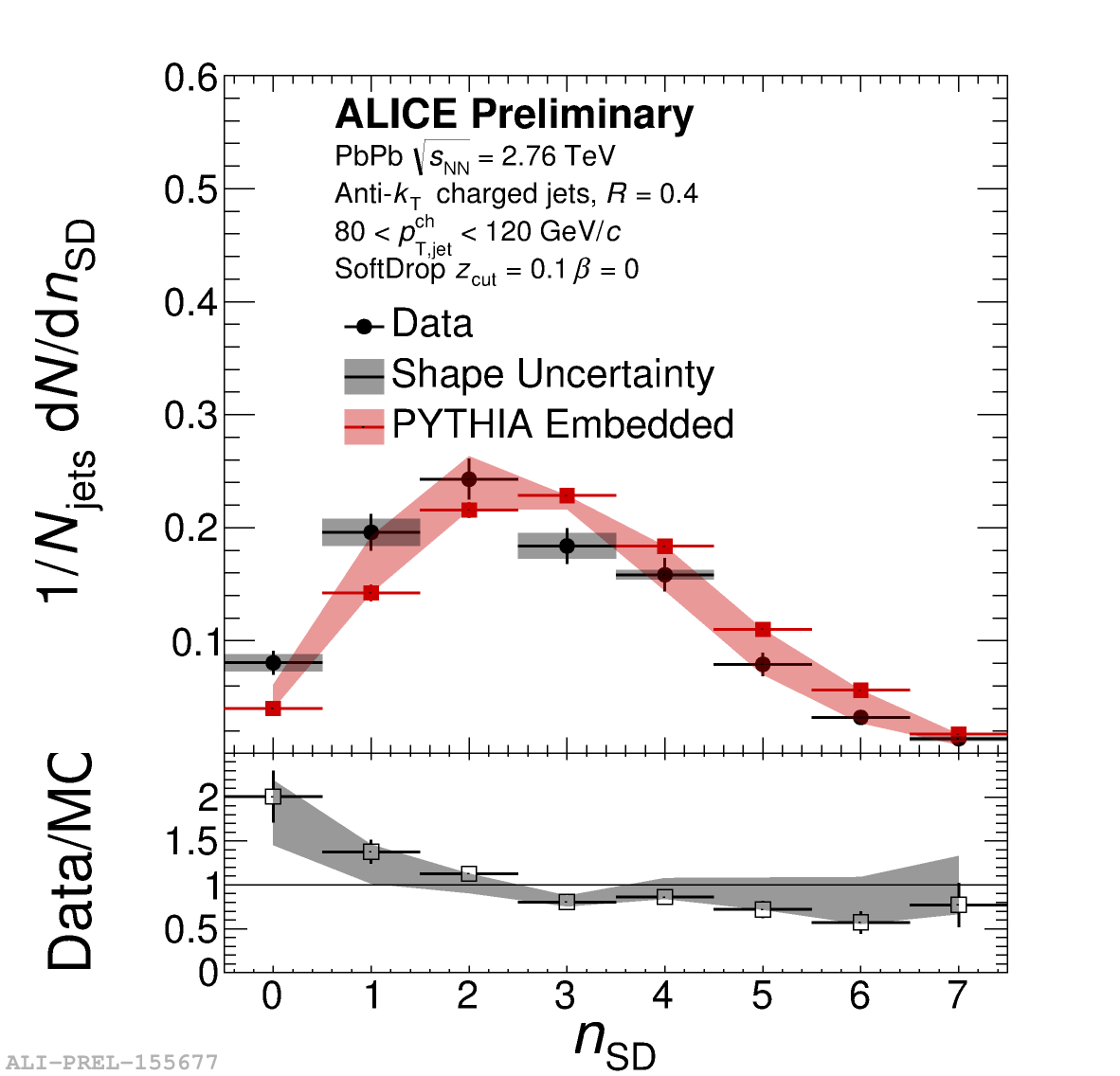} 
\captionof{figure}{ Number of Soft Drop branches in the PbPb data jets and in PYTHIA embedded jets.}
\label{fig:RecursiveDataRatioSD}
\end{minipage}
\begin{minipage}{.05\textwidth}
\includegraphics[width=0.8\textwidth]{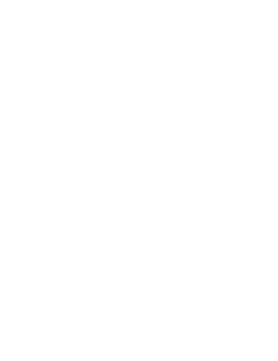} 
\centering
\end{minipage}
\begin{minipage}{.45\textwidth}
\centering
\includegraphics[width=0.9\textwidth]{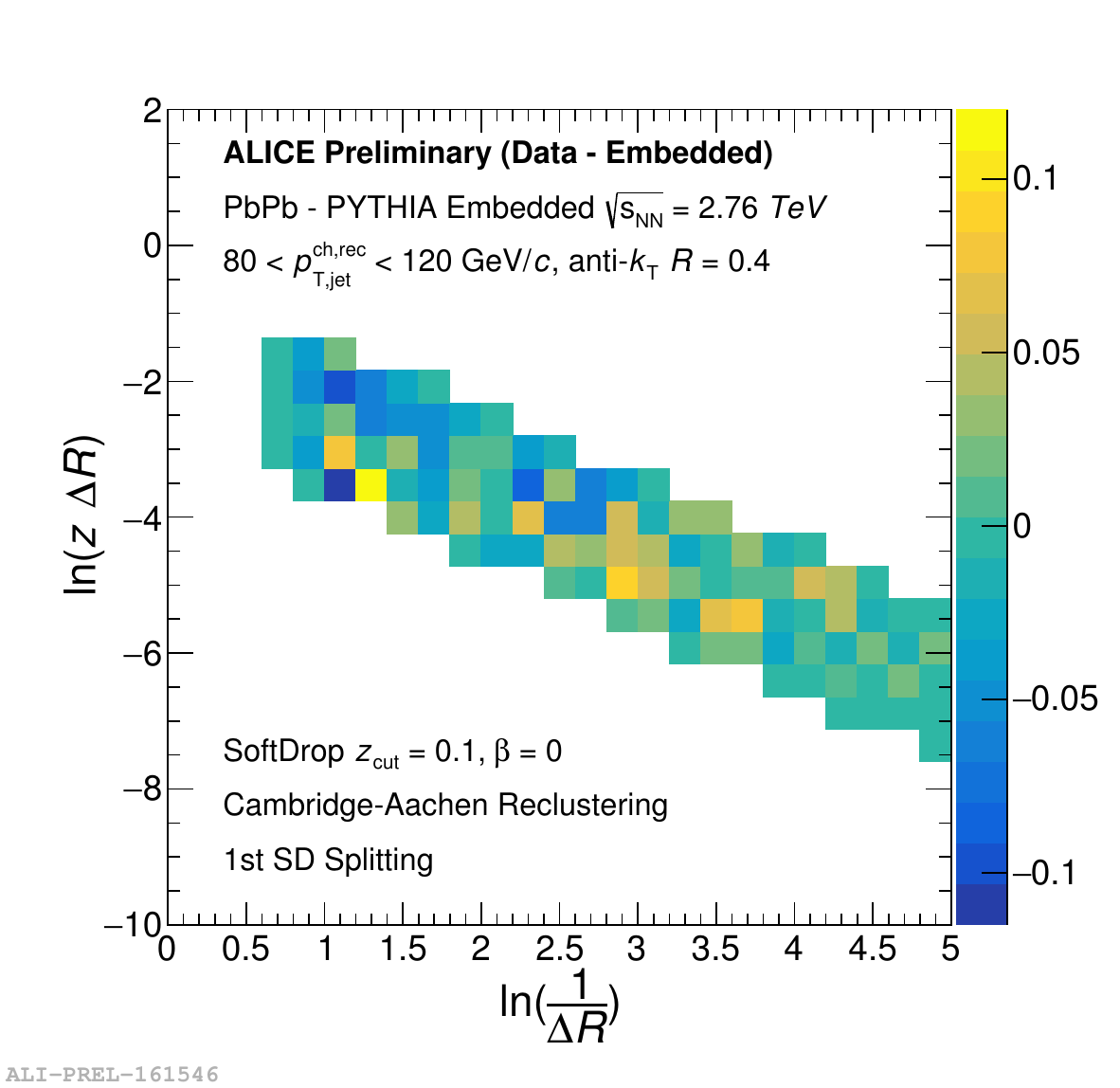}
\captionof{figure}{ Lund plane representing the difference between the first splitting identified in data and embedded PYTHIA.}
\label{fig:LundDiff}
\end{minipage}
\end{figure}

The measurement of $n_{SD}$ which corresponds to the number of branches in the jet clustering history that satisfy the Soft Drop cut is shown in Fig.~\ref{fig:RecursiveDataRatioSD}. This distribution shows no significant modification in the Pb--Pb data relative to vacuum data in the range 1$ \leq n_{\rm{SD}} \leq $ 7 with an enhancement observed in the ''untagged'' bin of \nsd = 0.  

Complimentary to the study of \zg ~differentially in \Rg , the Lund plane can be used to identify regions of significant enhancement or suppression. Fig.~\ref{fig:LundDiff} shows the Lund plane for the difference between the first splitting of jets in data and embedded PYTHIA. Using this representation of the phase space purely as a pictorial representation it is shown that in the collinear limit there is a general region of very slight enhancement whereas in the large angle limit there is an overall depletion in data. 

\section{Conclusions and Outlook}

Measurements of \zg~ and \nsd~ have been reported in pp and Pb--Pb collisions at \sqrts = 7 TeV and \sqrtsNN = 2.76 TeV, respectively. A good agreement between pp collisions and PYTHIA is observed and PYTHIA is used as a vacuum reference for comparison to Pb--Pb collisions. In Pb--Pb collisions a significant modification of the \zg~ distribution is observed at large angles whilst in the collinear limit the distribution is consistent across the full range. The modification in the large angle limit is attributed to a suppression of the symmetric splittings by normalising the distributions to the total number of jets in the jet transverse momentum range. The number of splittings, \nsd, is unmodified in Pb--Pb collisions suggesting that the modification of the \zg~ distribution isn't driven by an enhancement of splittings in any particular region of phase space. 

%% The Appendices part is started with the command \appendix;
%% appendix sections are then done as normal sections
%% \appendix

%% \section{}
%% \label{}

%% References
%%
%% Following citation commands can be used in the body text:
%% Usage of \cite is as follows:
%%   \cite{key}         ==>>  [#]
%%   \cite[chap. 2]{key} ==>> [#, chap. 2]
%%

%% References with BibTeX database:

\bibliographystyle{elsarticle-num}
\bibliography{QM18Proceedings_HarryAndrews_v4.bib}

%% Authors are advised to use a BibTeX database file for their reference list.
%% The provided style file elsarticle-num.bst formats references in the required Procedia style

%% For references without a BibTeX database:

% \begin{thebibliography}{00}

%% \bibitem must have the following form:
%%   \bibitem{key}...
%%

% \bibitem{}

% \end{thebibliography}

\end{document}

%% file: Include/Include.tex
\newcommand{\CommentBlock}[1]{}

\newcommand{\sqrtsNN}{\ensuremath{\sqrt{s_{\mathrm {NN}}}}}
\newcommand{\sqrts}{\ensuremath{\sqrt{s}}}

\newcommand{\zg}{\ensuremath{z_{\mathrm{g}}}}
\newcommand{\Rg}{\ensuremath{R_{\mathrm{g}}}}
\newcommand{\nsd}{\ensuremath{n_{\mathrm{SD}}}}

\newcommand{\pTjetch}{\ensuremath{p_\mathrm{T,jet}^\mathrm{ch}}}
\newcommand{\pTjetchrec}{\ensuremath{p_\mathrm{T,jet}^\mathrm{ch,rec}}}

\newcommand{\DeltaR}{\ensuremath{\Delta{R}}}